\documentclass[twoside,11pt]{article}
\usepackage{amsfonts}
\usepackage{mathrsfs}
\usepackage[utf8]{inputenc} 
\usepackage{float}
\usepackage{enumitem}
\usepackage{amsmath,amssymb}
\usepackage{latexsym}
\usepackage{enumitem} 
\usepackage{longtable}
\usepackage{booktabs}
\usepackage{epstopdf}
\usepackage{subcaption}
\usepackage{tabularx}
\usepackage{tikz}
\usetikzlibrary{arrows.meta,calc,fit,positioning}
\usepackage{geometry}  
\usepackage[active]{srcltx}
\usepackage{graphicx}
\usepackage{booktabs}
\usepackage{multirow}
\usepackage{siunitx}
\usepackage{xurl} 
\usepackage{amsmath}
\usepackage{setspace}
\usepackage{graphicx} 
\usepackage{algpseudocode}
\usepackage{algorithm}

\usepackage{subcaption} 
\usepackage[
    bookmarks=true,         
    bookmarksnumbered=true, 
    colorlinks=true, pdfstartview=FitV, linkcolor=blue, citecolor=blue,
    urlcolor=blue]{hyperref}

\usepackage{fancyhdr}
\usepackage{lipsum} 
\usepackage{listings} 
\usepackage{xcolor} 
\definecolor{codegreen}{rgb}{0,0.6,0}
\definecolor{codegray}{rgb}{0.5,0.5,0.5}
\definecolor{codepurple}{rgb}{0.58,0,0.82}
\definecolor{backcolour}{rgb}{0.95,0.95,0.92}

\lstdefinestyle{mystyle}{
    backgroundcolor=\color{backcolour},
    commentstyle=\color{codegreen},
    keywordstyle=\color{magenta},
    numberstyle=\tiny\color{codegray},
    stringstyle=\color{codepurple},
    basicstyle=\footnotesize\ttfamily,
    breakatwhitespace=false,
    breaklines=true,
    captionpos=b,
    keepspaces=true,
    numbers=left,
    numbersep=5pt,
    showspaces=false,
    showstringspaces=false,
    showtabs=false,
    tabsize=2
}

\providecommand{\U}[1]{\protect\rule{.1in}{.1in}}

\usepackage{listings}
\usepackage{titlesec}
\titleformat{\section}
{\normalfont\Large\bfseries}{\thesection.}{1em}{}
\usepackage{subcaption}
\renewenvironment{enumerate}{
  \begin{list}{}{
    \setlength{\labelwidth}{0pt}
    \setlength{\leftmargin}{0pt}
    \setlength{\itemindent}{-2em}
    \setlength{\itemsep}{0.5em}
    \setlength{\parsep}{0pt}
    \setlength{\topsep}{0pt}
    \setlength{\partopsep}{0pt}
    \setlength{\listparindent}{2em}
    \setlength{\labelsep}{0pt}
    \setlength{\itemindent}{0pt}
    \setlength{\leftmargin}{2em}
    \setlength{\rightmargin}{0pt}
  }
}{
  \end{list}
}

\begin{document}
\date{}
\title{\Large \textbf{A2 Copula-Driven Spatial Bayesian Neural Network For Modeling Non-Gaussian Dependence: A Simulation Study}}
\vspace{1ex}
\author{Agnideep Aich${ }^{1}$\thanks{Corresponding author: Agnideep Aich, \texttt{agnideep.aich1@louisiana.edu}, ORCID: \href{https://orcid.org/0000-0003-4432-1140}{0000-0003-4432-1140}}
 \hspace{0pt}, Sameera Hewage${ }^{2}$ \hspace{0pt}, Md Monzur Murshed${ }^{3}$ \hspace{0pt}, Ashit Baran Aich${ }^{4}$ \hspace{0pt}, \\Amanda Mayeaux${ }^{5}$ \hspace{0pt}, Asim K. Dey${ }^{6}$ \hspace{0pt}, Kumer P. Das${ }^{7}$ \hspace{0pt} and Bruce Wade${ }^{8}$ \\
\\ ${ }^{1}$ ${ }^{8}$ Department of Mathematics, University of Louisiana at Lafayette, \\ Lafayette, Louisiana, USA \\ ${ }^{2}$ Department of Physical Sciences \& Mathematics, West Liberty University, \\ West Liberty, West Virginia, USA \\ ${ }^{3}$ Department of Mathematics and Statistics, Minnesota State University, \\ Mankato, Minnesota, USA\\ ${ }^{4}$ Department of Statistics, formerly of Presidency College, \\ Kolkata, India \\ ${ }^{5}$ Department of Kinesiology, University of Louisiana at Lafayette, \\Lafayette, Louisiana, USA \\${ }^{6}$ Department of Mathematics and Statistics, Texas Tech University, \\ Lubbock, Texas, USA \\ ${ }^{7}$ The Office of Vice President for Research, Innovation, and Economic Development,\\ University of Louisiana at Lafayette, Lafayette, Louisiana, USA\\}
\date{}
\maketitle
\vspace{-20pt}
\begin{abstract}
\noindent
In this paper, we introduce the \textbf{A2 Copula Spatial Bayesian Neural Network (A2-SBNN)}, a predictive spatial model designed to map coordinates to continuous fields while capturing both typical spatial patterns and extreme dependencies. By embedding the dual-tail novel Archimedean copula viz. A2 directly into the network’s weight initialization, A2-SBNN naturally models complex spatial relationships, including rare co-movements in the data. The model is trained through a calibration-driven process combining Wasserstein loss, moment matching, and correlation penalties to refine predictions and manage uncertainty. Simulation results show that A2-SBNN consistently delivers high accuracy across a wide range of dependency strengths, offering a new, effective solution for spatial data modeling beyond traditional Gaussian-based approaches.
\end{abstract}

\noindent \textbf{Keywords:} A2 Copula, A2-SBNN, Spatial Bayesian Neural Network, Non-Gaussian Dependence, Tail Dependency, Archimedean Copula

\medskip
\noindent \textit{MSC 2020 Subject Classification:} 62H12, 62P10, 65C20, 62F15, 68T07

\section{Introduction}

Spatial data analysis plays a critical role in a wide range of scientific fields, including environmental sciences, epidemiology, and finance. Traditional statistical methods for spatial modeling often rely on Gaussian assumptions, which offer mathematical convenience but fall short when faced with the complexities of real world data. Many spatial datasets display heavy tailed behavior, skewness, and asymmetric dependencies, characteristics that Gaussian-based models struggle to handle effectively.

\medskip \noindent To overcome these limitations, modern approaches have turned to more flexible tools that separate marginal distributions from joint dependency structures. Copulas, originally introduced by Sklar (1959) and extensively developed by \href{https://doi.org/10.1007/0-387-28678-0}{Nelsen (2006)}, provide a robust framework for modeling complex, non-Gaussian dependencies. Among these, Archimedean copulas are particularly appealing due to their analytical simplicity and adaptability. Building on this foundation, \href{http://dx.doi.org/10.1080/03610926.2024.2440577}{Aich et al. (2025)} introduced two new Archimedean copulas viz. A and B. These are renamed as A1 and A2 in this article, and these were specifically designed to capture dual tail dependencies which is an essential feature when modeling extreme co-movements in spatial data.

\medskip \noindent In this work, we present what is, to the best of our knowledge, the first spatial neural network that directly integrates a copula-driven dependency structure into its architecture through weight initialization. We introduce the \textit{A2 Copula Spatial Bayesian Neural Network (A2-SBNN)}, a novel framework that uses the A2 copula to embed complex spatial dependencies directly into the model from the very beginning of training. By initializing the weights of the neural network using the inverse generator of the A2 copula, the model is inherently equipped to handle both upper and lower tail dependencies, ensuring that extreme spatial events and localized interactions are effectively captured.

\medskip \noindent At its core, the A2-SBNN is a predictive spatial model designed to map spatial coordinates to a continuous spatial field while preserving the intricate dependency structures that define real world spatial data. Unlike conventional spatial models, which tend to prioritize smooth central tendencies, the A2-SBNN is built to respect both the average behavior of the data and its most extreme variations. This makes the model particularly valuable in applications where understanding the full range of spatial variability including rare or extreme events is essential.

\medskip \noindent To calibrate the A2-SBNN, we introduce a custom loss function that goes beyond traditional mean squared error (MSE). Our approach incorporates the Wasserstein distance, moment matching, and correlation penalties, which together help align the model’s predictions with the target spatial field while ensuring that the dependency structure remains accurate. We rigorously evaluate the model across a wide range of tail dependence intensities, controlled by the parameter $\theta$ testing the network's performance from moderate to extreme spatial behaviors.

\medskip \noindent This integration of the A2 copula within a spatial Bayesian neural network represents a significant step forward in spatial modeling. Previous neural network-based approaches have largely depended on Gaussian assumptions, limiting their usefulness when faced with non-Gaussian, heavy tailed spatial data. In contrast, A2-SBNN bridges the gap between deep learning and advanced dependency modeling, offering a framework that is capable of handling the full complexity of spatial phenomena.

\medskip \noindent Our simulation studies demonstrate that the A2-SBNN achieves strong predictive performance across a broad range of dependency structures, with consistently high correlations and low RMSE values, even as tail behavior becomes more pronounced. Residual analysis confirms that the model maintains well-calibrated predictions, with residuals remaining normal across all tested scenarios, despite varying levels of tail dependency.

\medskip \noindent In summary, this work introduces the A2-SBNN as a powerful new tool for spatial prediction in settings where traditional models fall short. By embedding the A2 copula into the neural network's architecture and applying a carefully designed calibration process, we provide a framework capable of accurately modeling non-Gaussian spatial fields, capturing both central trends and extreme behaviors. The remainder of this paper is organized as follows. Section 2 reviews related work. Section 3 introduces the A2 copula and details its integration into the neural network architecture. Section 4 presents simulation studies evaluating the performance of A2-SBNN across a wide range of tail dependencies and also reports the residual analysis. Finally, Section 5 concludes with a discussion of the findings and outlines directions for future research.

\section{Background and Motivation}

Traditional methods in spatial statistics have long relied on Gaussian processes and related techniques (\href{https://doi.org/10.1002/9781119115151}{Cressie, 1993}; \href{https://doi.org/10.7551/mitpress/3206.001.0001}{Rasmussen and Williams, 2005}). These models are popular because they offer elegant mathematical properties and make inference relatively straightforward. However, they come with an important limitation: they assume the underlying spatial dependence is Gaussian. When spatial data exhibit heavy tails, skewness, or asymmetric dependencies, Gaussian-based models often fall short, struggling to capture the full complexity of the underlying relationships. To address these gaps, researchers have developed alternative stochastic process models tailored for spatial extremes. Notable contributions include the framework by \href{https://doi.org/10.1214/11-STS376}{Davison et al. (2012)} for modeling spatial extremes, and the comprehensive review by \href{https://doi.org/10.1002/wics.1537}{Huser and Wadsworth (2020)}. While these models offer greater flexibility, they often become computationally intensive, especially with large, high-dimensional spatial datasets.

\medskip
\noindent More recently, deep learning has started to make an impact in spatial statistics, offering new ways to address these challenges. Bayesian neural networks (BNNs), for example, have been adapted for spatial tasks by incorporating spatial embeddings and allowing for spatially varying parameters. One example is the work by \href{https://doi.org/10.1016/j.spasta.2024.100825}{Zammit-Mangion et al. (2024)}, which introduced the concept of \textit{Spatial Bayesian Neural Networks} (SBNNs). These models are specifically designed to learn spatial patterns directly from the data, reducing the need for manual tuning and improving generalization across different types of spatial problems.

\medskip
\noindent At the same time, copula theory has gained popularity as a flexible tool for capturing dependencies in multivariate data. Copulas allow us to separate the modeling of marginal distributions from the dependency structure itself, which makes them particularly useful in fields like finance, healthcare, and environmental science. Among the different copula families, Archimedean copulas stand out for their simplicity and adaptability to various forms of dependency. Recently, two new Archimedean copulas first introduced as the A and B copulas by \href{http://dx.doi.org/10.1080/03610926.2024.2440577}{Aich et al. (2025)} and now renamed as A1 and A2 were proposed specifically to handle dual tail dependencies. Although traditional likelihood-based estimation is difficult with these copulas because of their complex density functions, alternative estimation methods have been developed that strike a balance between theory and computational efficiency.

\medskip
\noindent In parallel with these advances, the Wasserstein distance has become a standard tool in training deep generative models. In particular, the work by \href{https://doi.org/10.48550/arXiv.1704.00028}{Gulrajani et al. (2017)} introduced a gradient penalty approach to enforce Lipschitz continuity, which has since become widely adopted in Wasserstein Generative Adversarial Networks (GANs). In our work, we apply this idea in the calibration phase of the A2-SBNN, combining the Wasserstein loss with moment matching and correlation penalties to ensure that the model’s output is well-aligned with the target distribution.

\medskip
\noindent Our approach brings these research directions together. By embedding the A2 copula into a spatial neural network, we created the A2 Copula Spatial Bayesian Neural Network (A2-SBNN). This framework is designed to model non-Gaussian spatial dependencies, particularly in situations where extreme values and asymmetric tail behaviors are important. Thanks to the A2 copula’s dual tail dependency, the model can simultaneously capture both upper and lower tail dependencies, making it especially useful for data where extreme co-movements occur in space. By combining the strengths of the A2 copula and Wasserstein-based calibration, the A2-SBNN goes beyond traditional spatial models, expanding what is possible in spatial deep learning. Our work is also related to DeepKriging (\href{https://doi.org/10.48550/arXiv.2007.11972}{Chen et al., 2020}), which uses spatial basis functions in neural networks but stays within a Gaussian framework.

\medskip
\noindent In short, while previous research has tackled the challenges of modeling complex spatial dependencies and capturing non-Gaussian behaviors separately, our A2-SBNN is the first to directly integrate a copula-based dependency structure with Wasserstein-based calibration inside a spatial neural network. This combination is designed to improve predictive accuracy, handle extreme values more effectively, and offer better uncertainty quantification for real world problems where heavy tails and complex spatial dependencies are common.

\section{Methodology}

This work presents the A2 Copula Spatial Bayesian Neural Network (A2-SBNN), a novel spatial modeling framework designed to capture complex dependencies, including extreme co-movements in both lower and upper tails. Central to this architecture is the A2 copula, which governs the weight initialization of all fully connected layers, embedding dual-tail dependency into the network from the outset. This initialization method departs from traditional uniform or Gaussian-based schemes, enabling the network to effectively model intricate spatial relationships.

\subsection{A2 Copula-Based Weight Initialization}

The inverse generator function of the A2 copula is defined as:
\begin{equation}
\phi_{A2}^{-1}(t; \theta) = \frac{2 + t^{1/\theta} - \sqrt{(2 + t^{1/\theta})^2 - 4}}{2}, \quad \theta \geq 1,
\end{equation}
where \(t \sim U(0,1)\) and \(\theta\) is a parameter controlling the intensity of tail dependence. It is to be noted that Equation (1) is the corrected version of Equation (7) in \href{http://dx.doi.org/10.1080/03610926.2024.2440577}{Aich et al. (2025)}. \medskip  \\To initialize the weights:

\begin{enumerate}
    \item \textbf{1. } Uniform random variables \(t\) are sampled and clamped within \([10^{-9}, 1 - 10^{-9}]\) to avoid numerical instabilities.
    \item \textbf{2. }These samples are transformed through \(\phi_{A2}^{-1}(t; \theta)\) and scaled by a factor of 4.0 to control the spread of the weights.
    \item \textbf{3. }The scaled values are standardized and mapped through a sigmoid function to constrain them to \((0,1)\).
    \item \textbf{4. }The outputs are then passed through the inverse cumulative distribution function of a standard normal distribution.
    \item \textbf{5. }Finally, the weights are clipped within \(\pm 0.25/\sqrt{\theta}\), with a small epsilon (\(1 \times 10^{-3}\)) added to prevent extreme values.
\end{enumerate}
This process ensures that the network's initial weights encode dual-tail dependencies, promoting the learning of complex spatial behaviors while maintaining numerical stability.

\begin{figure}[H]
    \centering
    \includegraphics[width=\linewidth]{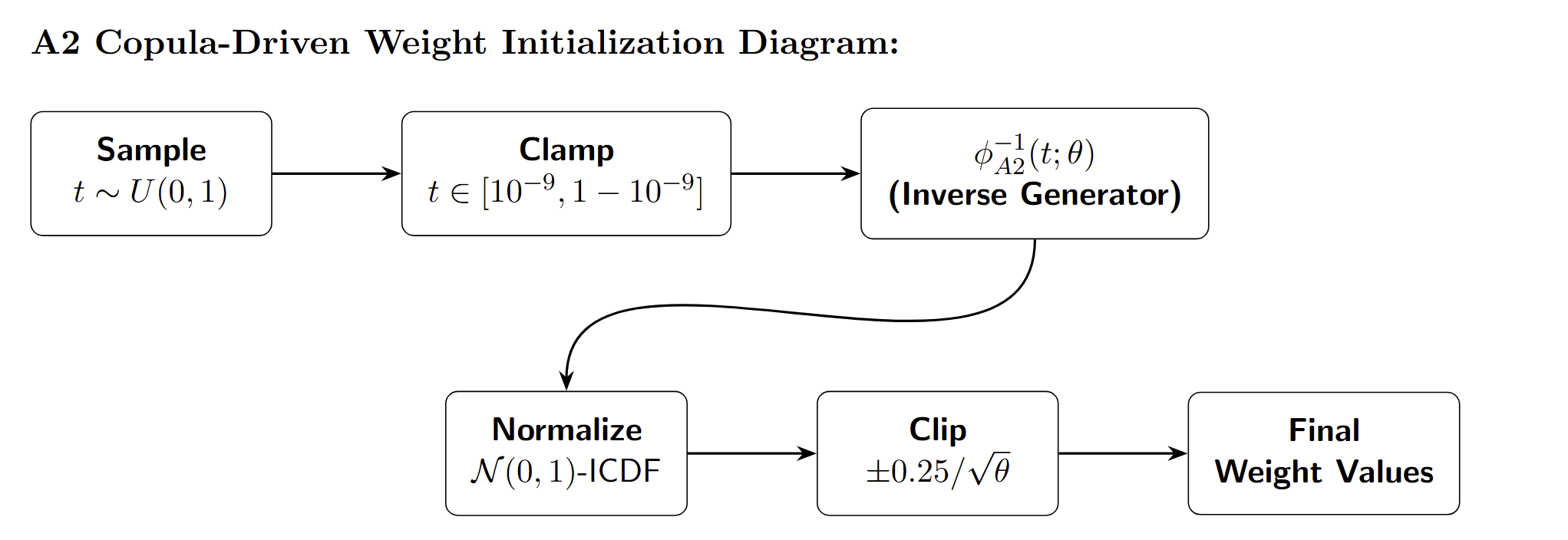}
    \caption{A2 Copula-Driven Weight Initialization Process}
    \label{fig:a2_initialization}
\end{figure}

\subsection{Fixed Target Field}

To provide a robust benchmark for calibration, a fixed spatial target field is synthesized using a squared-exponential covariance kernel computed from pairwise Euclidean distances on a spatial grid. The covariance matrix is decomposed via Cholesky decomposition, and noise from a Student's \(t\)-distribution is injected, producing a field with heavy tailed variability that mirrors realistic spatial phenomena.

\subsection{Bayesian Perspective}
In this framework, the A2 copula-based weight initialization acts as a Bayesian prior by embedding dual-tail dependencies into the network's weights. Additionally, by generating multiple forward passes during inference, the model approximates the predictive distribution, thereby enabling uncertainty quantification.

\subsection{A2-SBNN Architecture}

The architecture begins with a spatial embedding layer that projects spatial coordinates into a higher-dimensional feature space using radial basis functions (RBFs). Given \(K\) centers distributed over the spatial domain, the embedding for a spatial coordinate \(x \in \mathbb{R}^2\) is computed as:
\begin{equation}
\varphi(x)_k = \exp\left(-\frac{\|x - c_k\|^2}{\tau^2}\right), \quad k = 1, \ldots, K,
\end{equation}
where \(c_k\) represents the \(k\)-th center and \(\tau\) is a tunable length-scale parameter. 

\medskip
\noindent The embedded features pass through three fully connected hidden layers equipped with batch normalization and ELU activations. A residual connection bridges the first and second hidden layers, supporting stable gradient propagation and improved feature extraction. The hidden layers operate as:
\begin{align}
h_1 &= \text{ELU}\Big(\text{BatchNorm}\big(W_1\,\varphi(X) + b_1\big)\Big), \\
h_2 &= \text{ELU}\Big(\text{BatchNorm}\big(W_2\,h_1 + b_2\big)\Big) + h_1, \\
h_3 &= \text{ELU}\Big(\text{BatchNorm}\big(W_3\,h_2 + b_3\big)\Big).
\end{align}

\medskip
\noindent
The output layer concatenates the transformed hidden features, \( h_3 \), with the original spatial embedding \( \varphi(X) \), and applies a final linear transformation to produce the predictions. Specifically, the weight matrix \( W_{\text{out}} \) combines both the learned high-level features and the original spatial information, while \( b_{\text{out}} \) denotes the associated bias term. This design ensures that the network preserves both the deep representations of spatial interactions and the raw spatial structure in its final prediction:
\begin{equation}
\hat{y} = W_{\text{out}} \begin{bmatrix} h_3 \\ \varphi(X) \end{bmatrix} + b_{\text{out}},
\end{equation}
where \( W_{\text{out}} \in \mathbb{R}^{1 \times (D_1 + D_2)} \) operates on the concatenated feature vector of dimension \( D_1 + D_2 \), and \( b_{\text{out}} \in \mathbb{R} \) is a scalar bias.

\medskip
\noindent
\textbf{Schematic Diagram of the A2-SBNN Architecture:}

\begin{figure}[H]
    \centering
    \includegraphics[width=\linewidth]{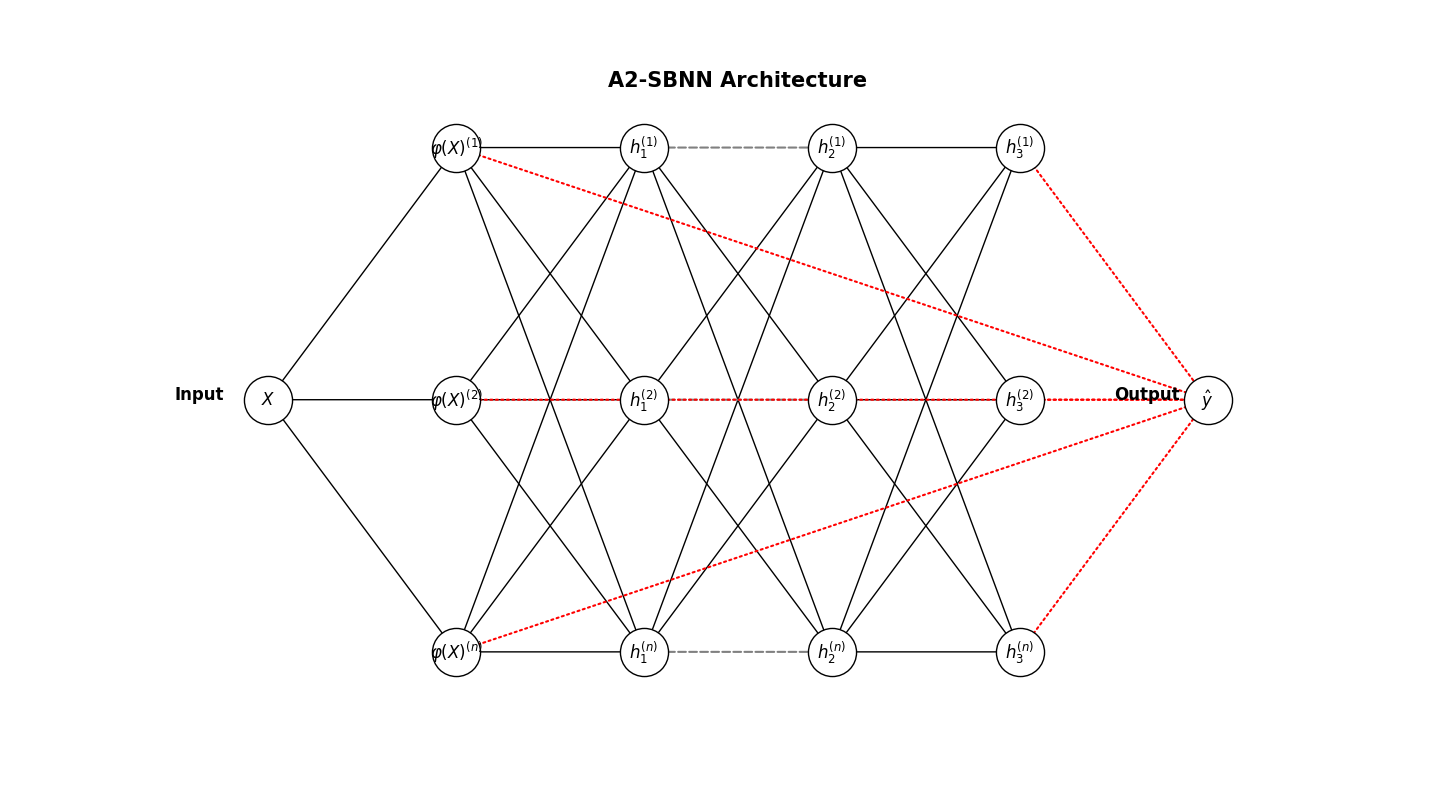}
    \caption{Schematic Diagram of the A2-SBNN Architecture}
    \label{fig:a2_sbnn_architecture}
\end{figure}

\medskip
\noindent
The input \(X\) represents the raw spatial coordinates. These are transformed through the embedding layer using radial basis functions (RBFs) to produce the embedded features:
\[
\varphi(X)^{(1)},\ \varphi(X)^{(2)},\ \ldots,\ \varphi(X)^{(n)}.
\]
Notably, this architecture follows a feed forward design, where each layer’s outputs feed into the next layer, supplemented by residual connections to enhance gradient flow.
The embedded features pass through a sequence of fully connected hidden layers. The first hidden layer produces outputs \(h_1^{(1)}, h_1^{(2)}, \ldots, h_1^{(n)}\), which are connected to the second hidden layer \(h_2^{(1)}, h_2^{(2)}, \ldots, h_2^{(n)}\). Residual (skip) connections, shown as dashed gray lines, link these two layers to improve gradient flow and stabilize training. The third hidden layer generates the final hidden outputs \(h_3^{(1)}, h_3^{(2)}, \ldots, h_3^{(n)}\).

\medskip
\noindent
The red dotted lines highlight the concatenation of the spatial embeddings \(\varphi(X)\) and the final hidden outputs \(h_3\) directly into the output layer. This combination preserves both localized spatial information and deeper learned features. The concatenated vector \([h_3, \varphi(X)]\) is then passed through a final linear transformation to produce the calibrated spatial prediction \(\hat{y}\).

\subsection{Model Calibration}

The A2-SBNN is trained via direct calibration against the fixed target field. The total calibration loss integrates:
\begin{equation}
L_{\text{cal}} = L_{\text{sup}} + \lambda_W L_W + \lambda_{\text{moment}} L_{\text{moment}} + \lambda_{\text{corr}} L_{\text{corr}},
\end{equation}
where \(L_{\text{sup}}\) is the mean squared error, \(L_W\) is the Wasserstein distance computed by a critic network with gradient penalty enforcement, \(L_{\text{moment}}\) matches means, and \(L_{\text{corr}}\) penalizes low Pearson correlation. Here, \(\lambda_W\) is a hyperparameter controlling the contribution of the Wasserstein distance \(L_W\) to the total calibration loss, balancing the trade-off between distributional alignment and predictive accuracy.

\subsection{Evaluation Metrics}

Model performance is evaluated through:
\begin{enumerate}

    \item \textbf{1. }Root Mean Squared Error (RMSE) for accuracy assessment.
    \item \textbf{2. }Pearson Correlation Coefficient to verify dependency preservation.
    \item \textbf{3. }Residual Analysis via histograms and the Shapiro-Wilk test for normality validation.
\end{enumerate}

\medskip
\noindent
In conclusion, the A2-SBNN framework introduces copula-driven initialization and direct calibration, offering a principled approach to spatial modeling that faithfully captures complex dependency structures and tail behaviors.


\section{Simulation Study}

In this section, we evaluate the performance of the A2 Copula Spatial Bayesian Neural Network (A2-SBNN) through direct calibration across a range of \(\theta\) values. We consider \(\theta \in \{1.5, 2.0, 3.0, 4.0, 5.0, 6.0, 7.0, 8.0, 9.0, 10.0\}\) to investigate how the strength of tail dependency affects the network's ability to capture complex spatial structures and extremes. For each \(\theta\), we present the calibrated predictions compared against the fixed target field.

\subsection{Calibrated Predictions}

\begin{figure}[H]
    \centering
    \begin{subfigure}[b]{0.3\textwidth}
        \includegraphics[width=\linewidth]{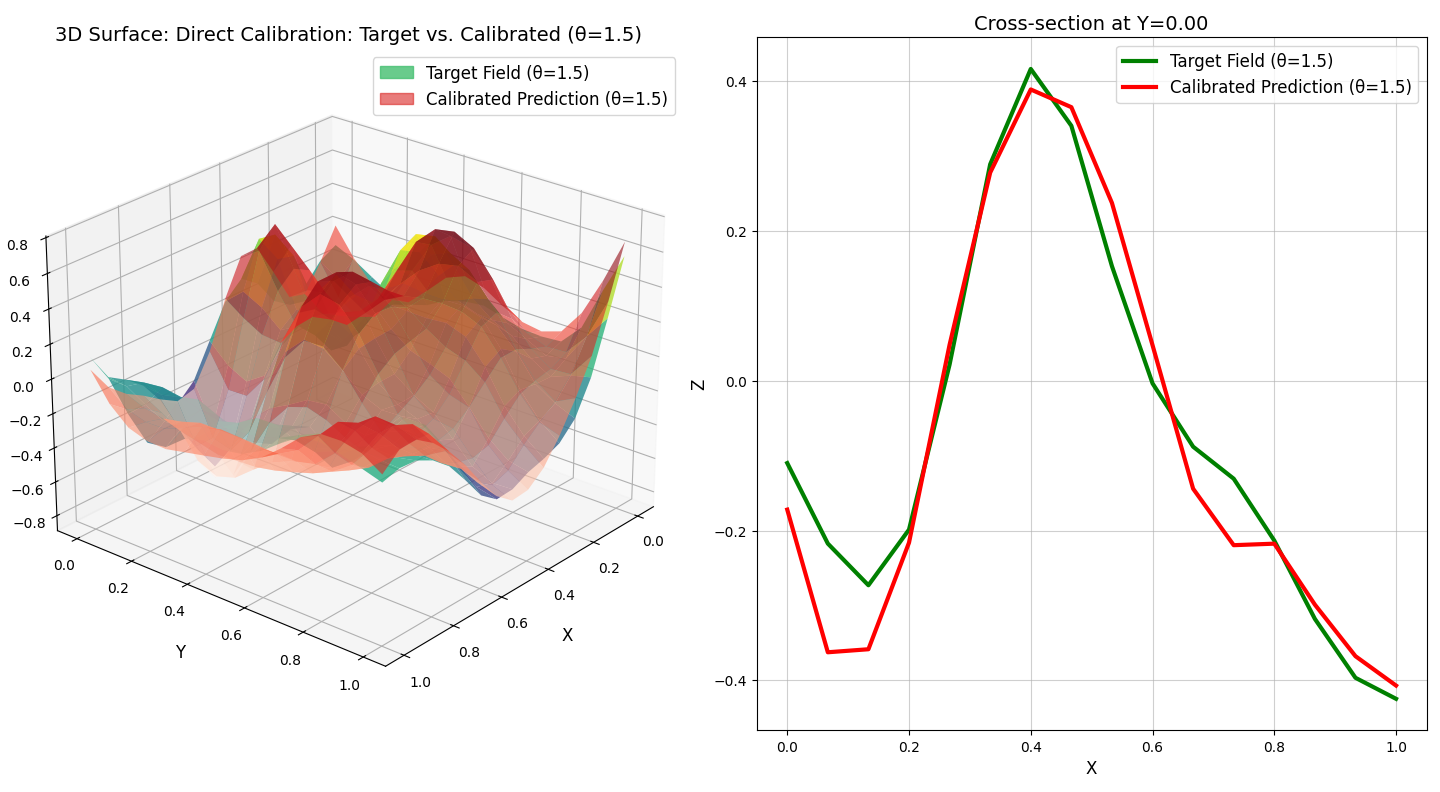}
        \caption{Fixed Target Field vs Calibrated Prediction for $\theta = 1.5$}
    \end{subfigure}
    \hfill
    \begin{subfigure}[b]{0.3\textwidth}
        \includegraphics[width=\linewidth]{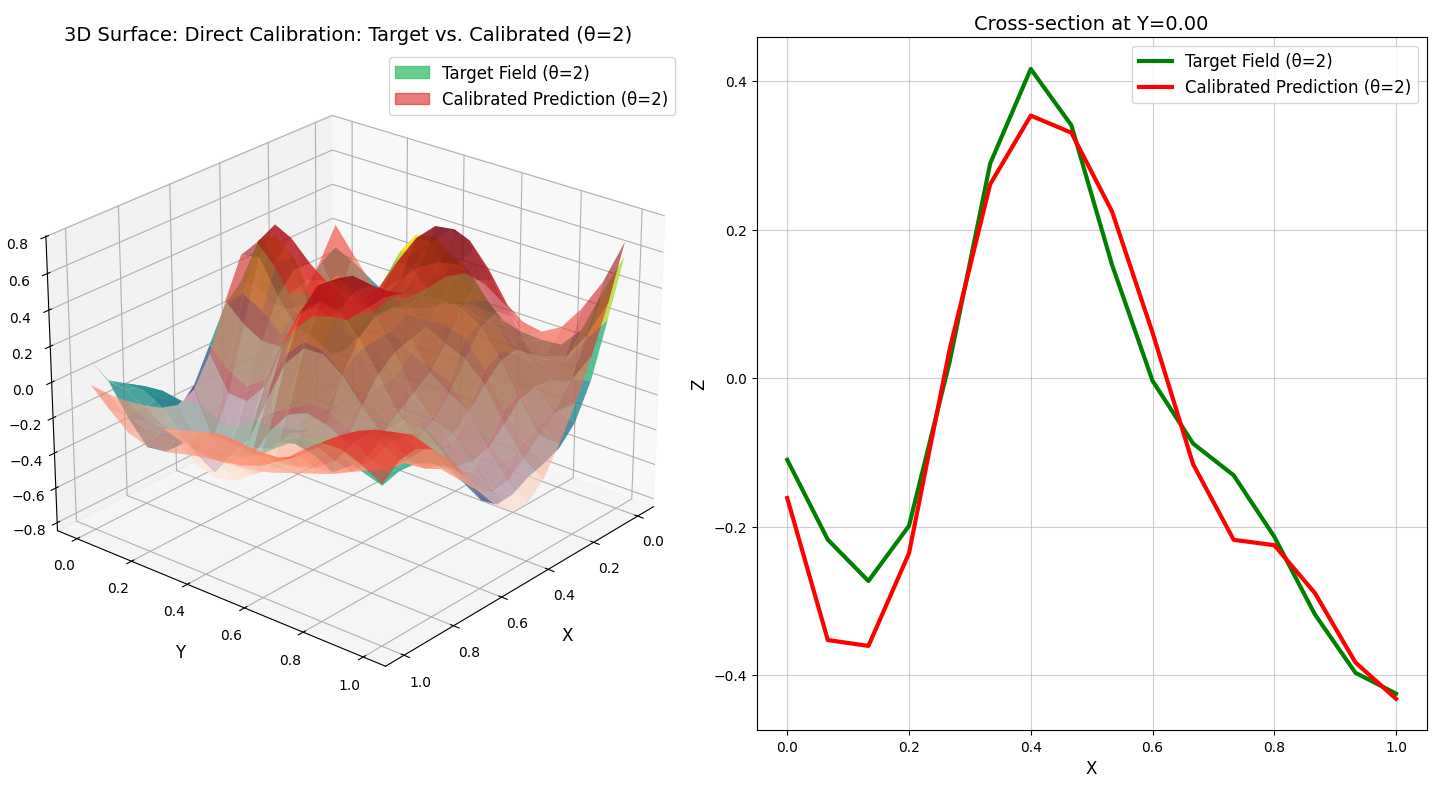}
        \caption{Fixed Target Field vs Calibrated Prediction for $\theta = 2.0$}
    \end{subfigure}
    \hfill
    \begin{subfigure}[b]{0.3\textwidth}
        \includegraphics[width=\linewidth]{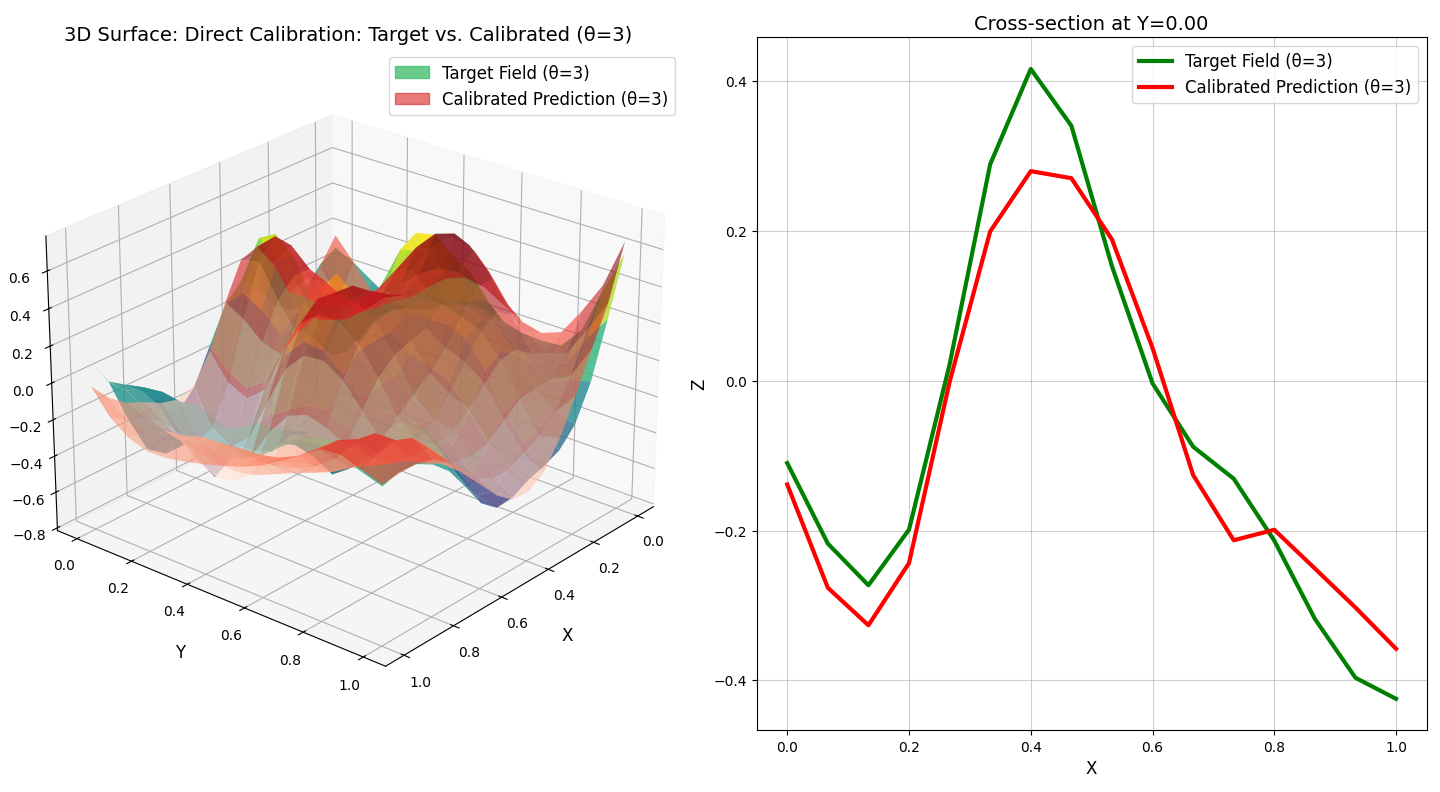}
        \caption{Fixed Target Field vs Calibrated Prediction for $\theta = 3.0$}
    \end{subfigure}

    \medskip

    \begin{subfigure}[b]{0.3\textwidth}
        \includegraphics[width=\linewidth]{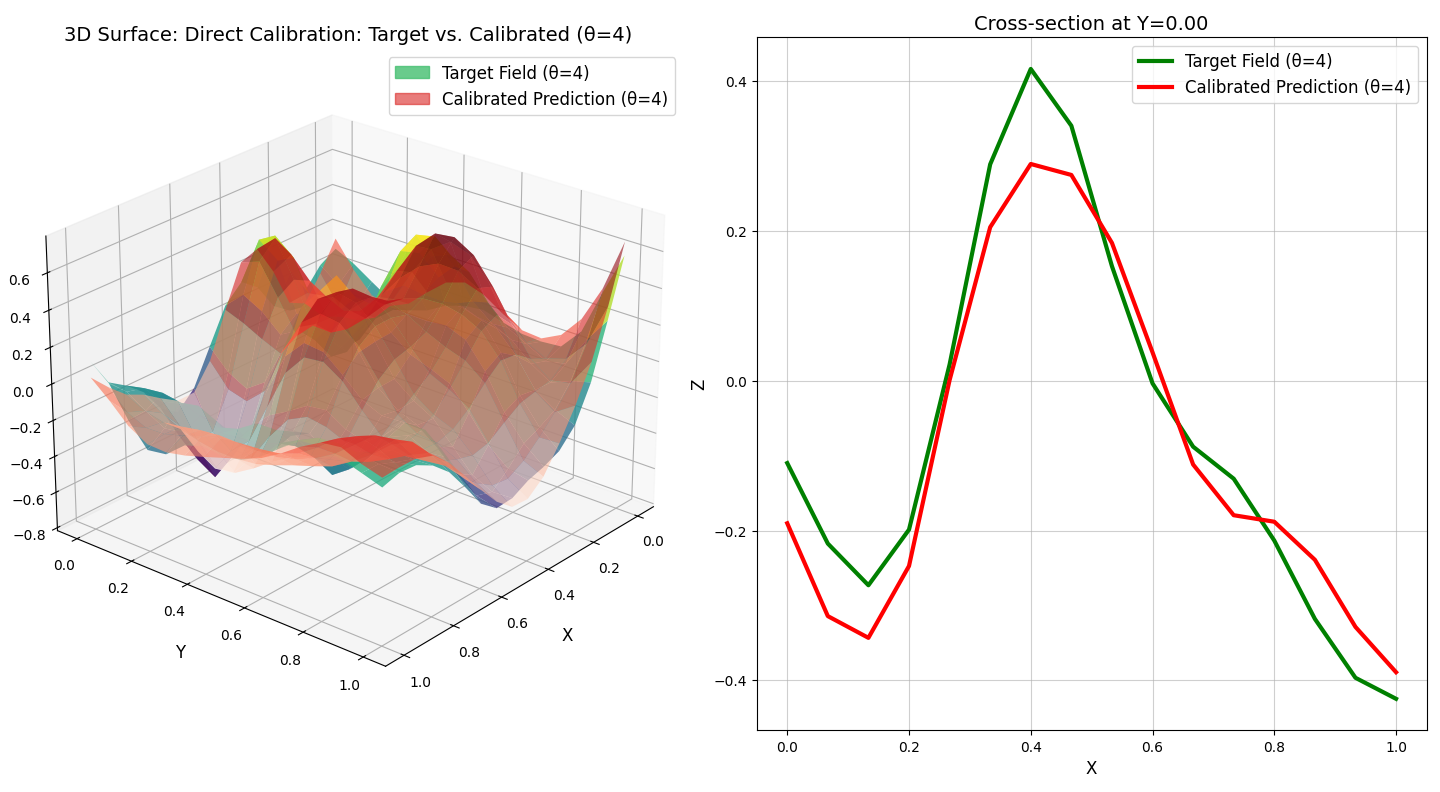}
        \caption{Fixed Target Field vs Calibrated Prediction for $\theta = 4.0$}
    \end{subfigure}
    \hfill
    \begin{subfigure}[b]{0.3\textwidth}
        \includegraphics[width=\linewidth]{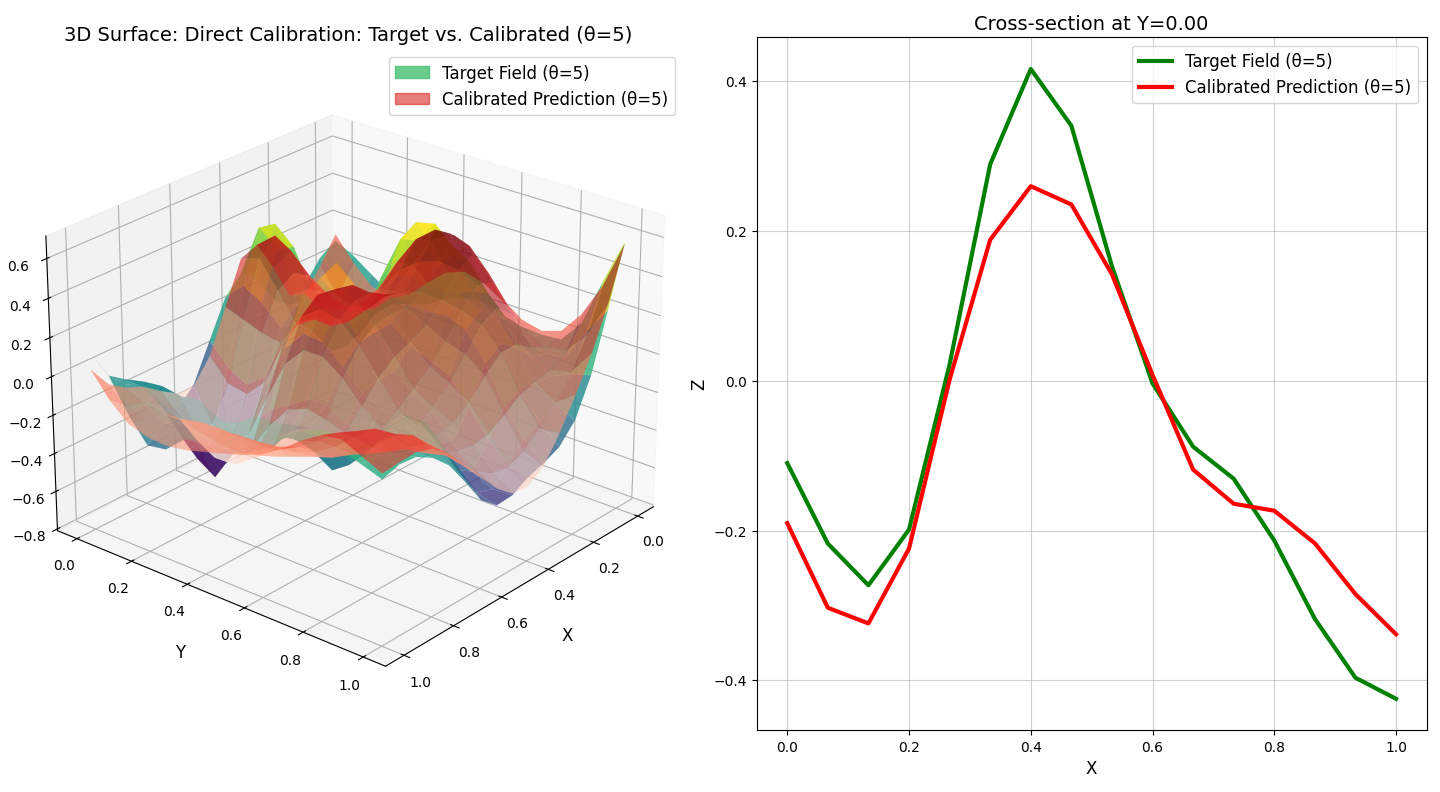}
        \caption{Fixed Target Field vs Calibrated Prediction for $\theta = 5.0$}
    \end{subfigure}
    \hfill
    \begin{subfigure}[b]{0.3\textwidth}
        \includegraphics[width=\linewidth]{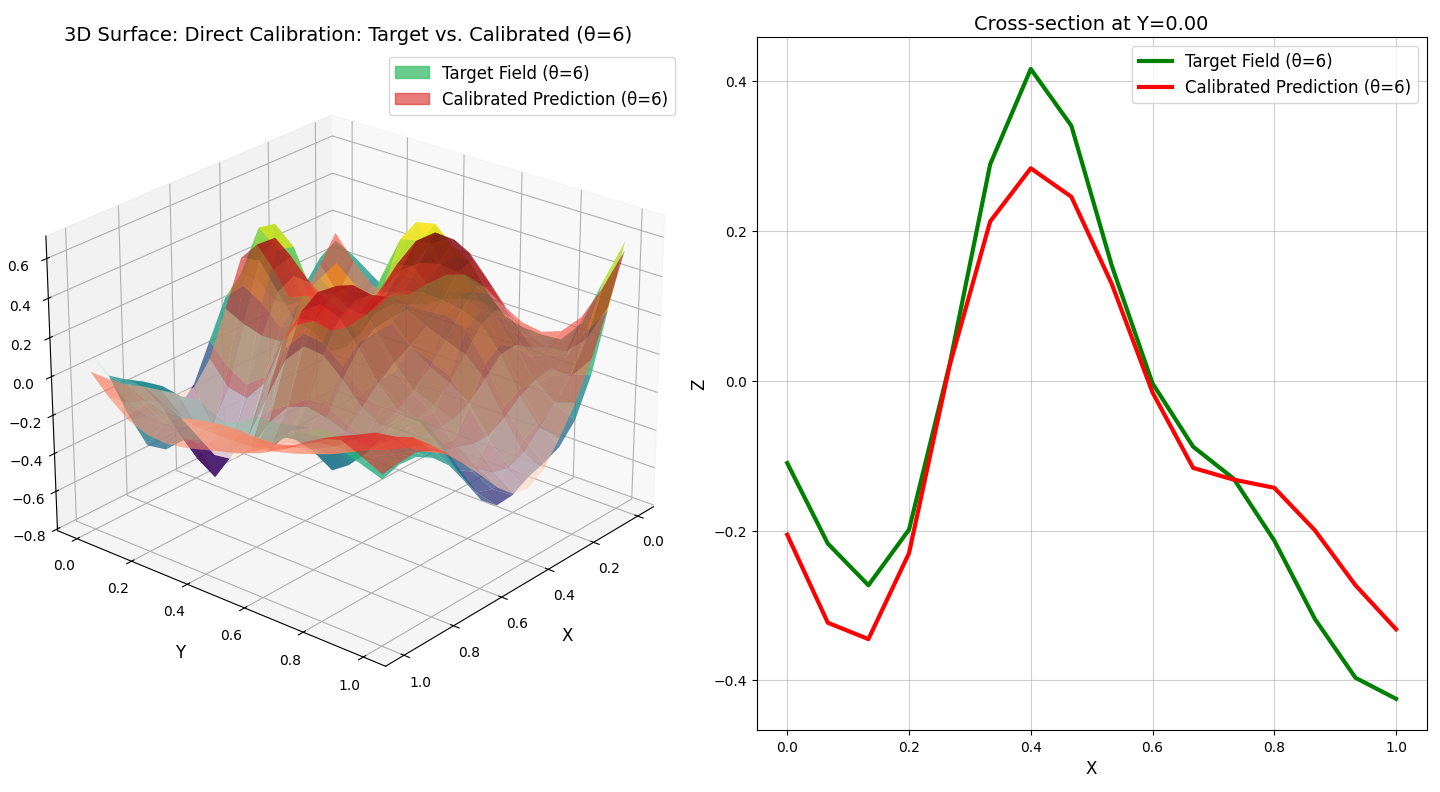}
        \caption{Fixed Target Field vs Calibrated Prediction for $\theta = 6.0$}
    \end{subfigure}

    \medskip

    \begin{subfigure}[b]{0.3\textwidth}
        \includegraphics[width=\linewidth]{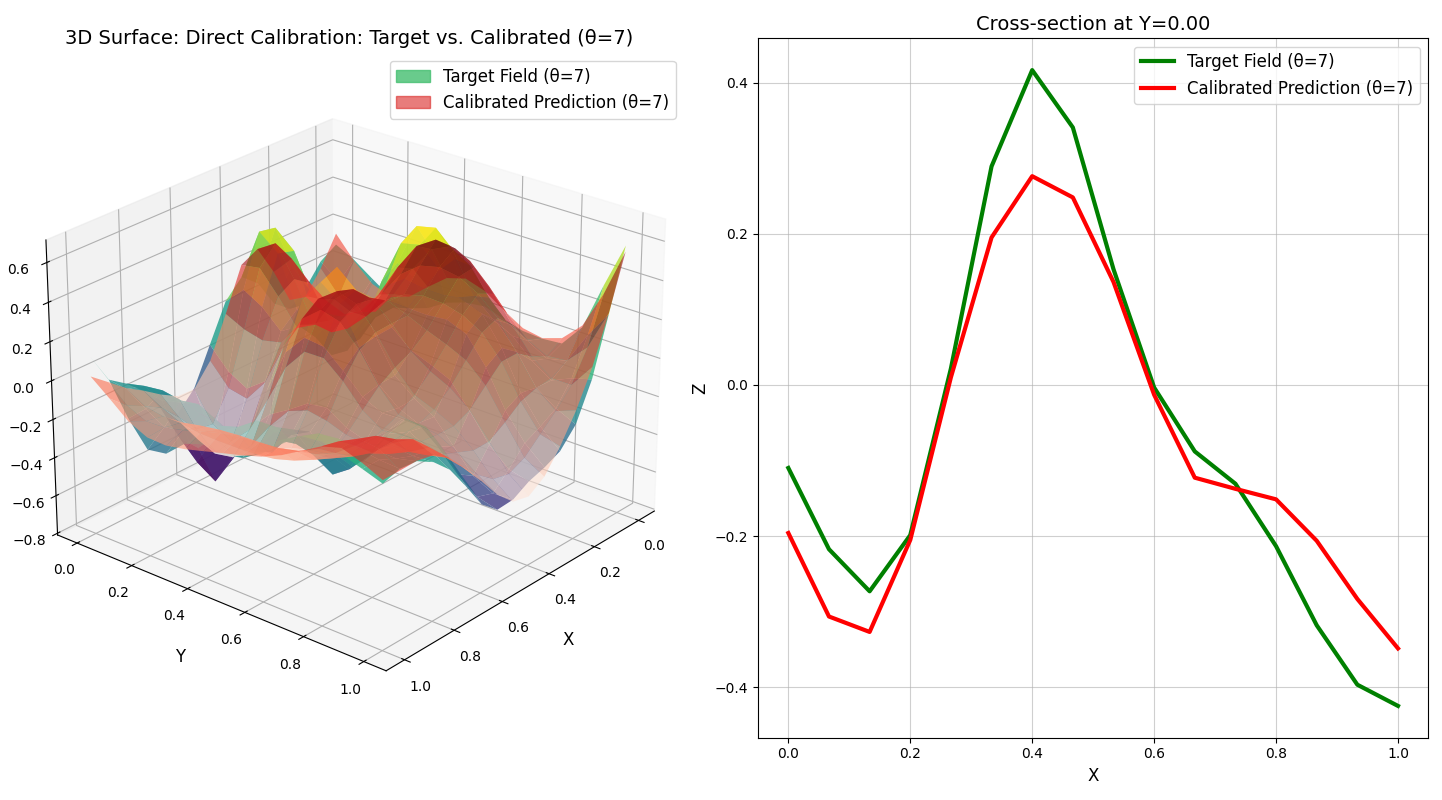}
        \caption{Fixed Target Field vs Calibrated Prediction for $\theta = 7.0$}
    \end{subfigure}
    \hfill
    \begin{subfigure}[b]{0.3\textwidth}
        \includegraphics[width=\linewidth]{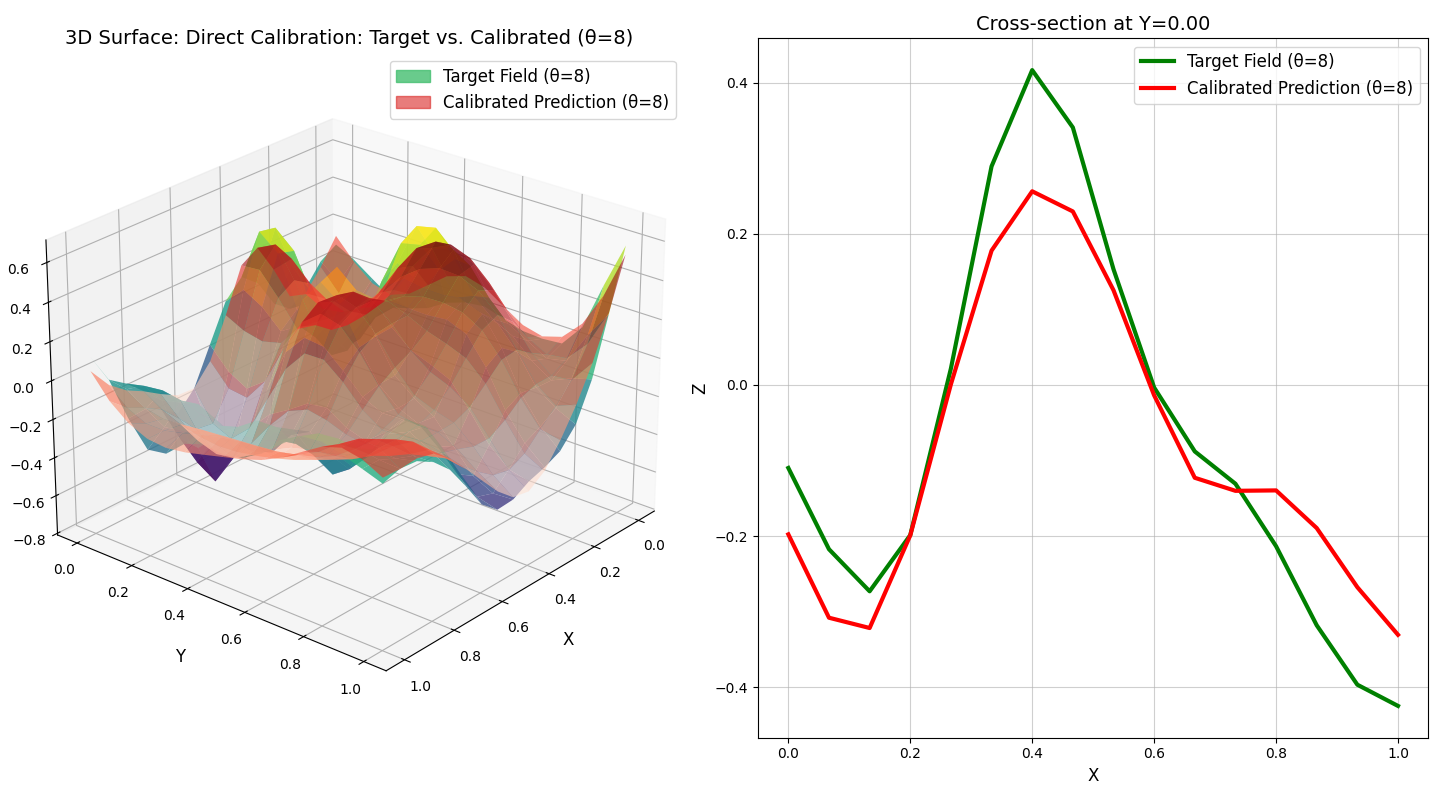}
        \caption{Fixed Target Field vs Calibrated Prediction for $\theta = 8.0$}
    \end{subfigure}
    \hfill
    \begin{subfigure}[b]{0.3\textwidth}
        \includegraphics[width=\linewidth]{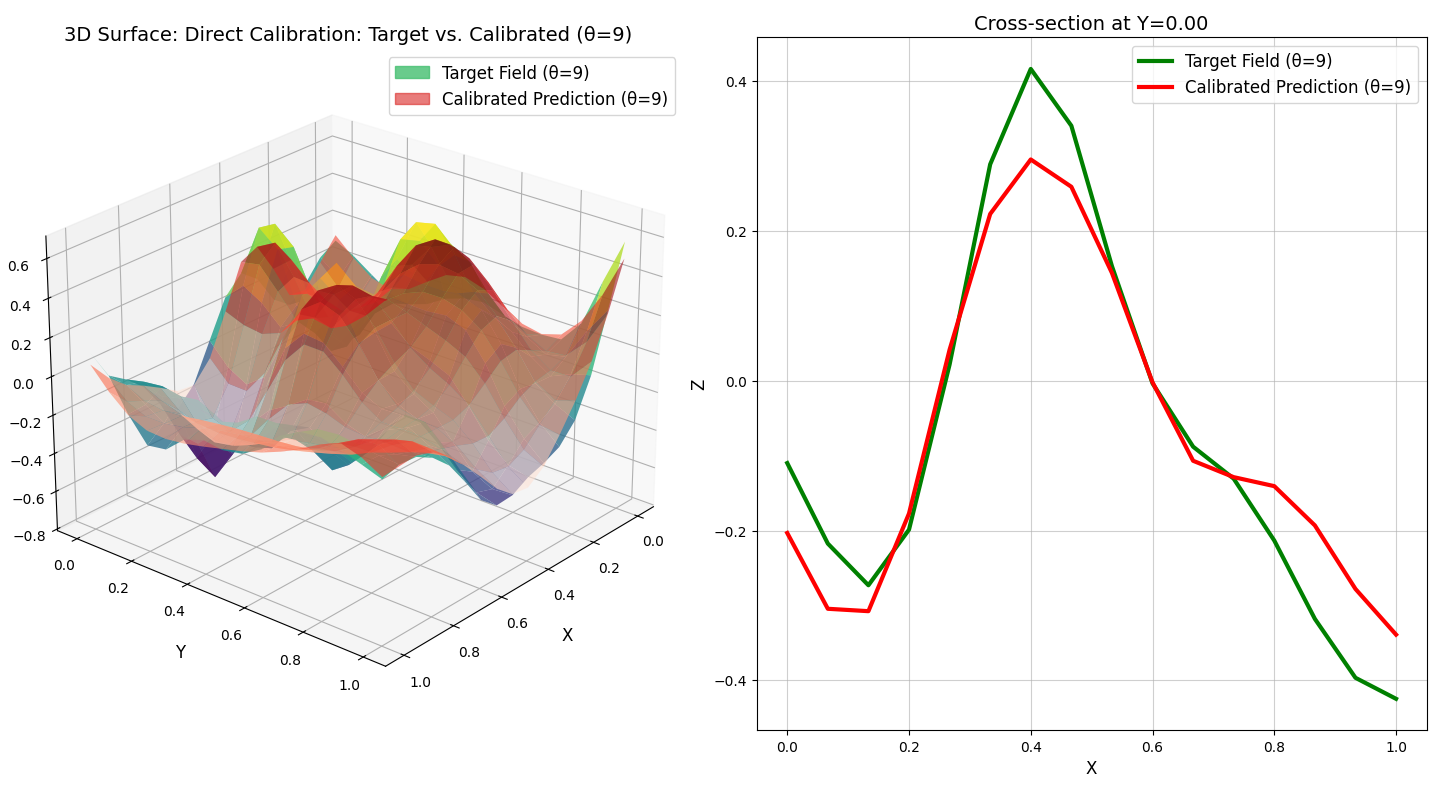}
        \caption{Fixed Target Field vs Calibrated Prediction for $\theta = 9.0$}
    \end{subfigure}

    \medskip

    \begin{subfigure}[b]{0.3\textwidth}
        \includegraphics[width=\linewidth]{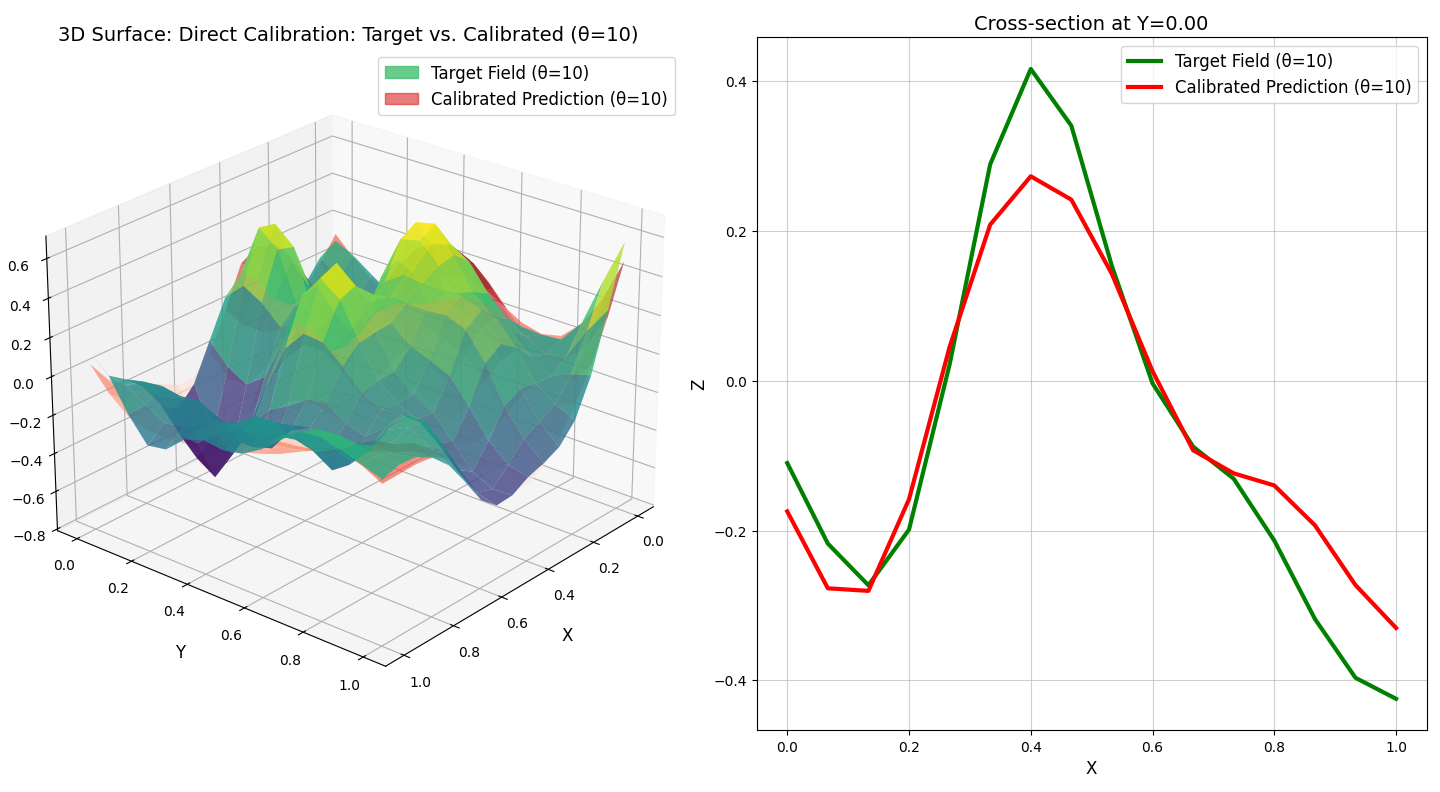}
        \caption{Fixed Target Field vs Calibrated Prediction for $\theta = 10.0$}
    \end{subfigure}

    \caption{Fixed Target Fields vs Calibrated Predictions for different $\theta$ values.}
    \label{fig:calibrated_results}
\end{figure}

For each value of \(\theta\), the model is directly calibrated to align with a fixed spatial target. Figure~\ref{fig:calibrated_results} displays the comparison between the target field and the calibrated prediction for all \(\theta\) values.

\noindent It is to be noted that The Target field is plotted with the ‘viridis’ colormap in Python and appears more green for $\theta$ = 10, while for other $\theta$ values, it may appear more red, purple or yellow due to the underlying data range and colormap transitions.

\subsubsection{Correlation and RMSE} 
The final correlation and RMSE between the calibrated prediction and the fixed target are shown below in Table~\ref{tab:calibration_results}.
\begin{table}[H]
    \centering
    \begin{tabular}{ccc}
        \toprule
        \(\theta\) & Correlation & RMSE \\
        \midrule
        1.5 & 0.9240 & 0.1071 \\
        2.0 & 0.9131 & 0.1152 \\
        3.0 & 0.8995 & 0.1088 \\
        4.0 & 0.9290 & 0.0888 \\
        5.0 & 0.9274 & 0.0833 \\
        6.0 & 0.9377 & 0.0764 \\
        7.0 & 0.9369 & 0.0769 \\
        8.0 & 0.9266 & 0.0825 \\
        9.0 & 0.9390 & 0.0758 \\
        10.0 & 0.9338 & 0.0799 \\
        \bottomrule
    \end{tabular}
    \caption{Calibration Performance Metrics of A2-SBNN across different \(\theta\) values.}
    \label{tab:calibration_results}
\end{table}

\noindent Table~\ref{tab:calibration_results} presents the calibration performance of the A2 Copula Spatial Bayesian Neural Network (A2-SBNN) across a range of \(\theta\) values from 1.5 to 10. The hyperparameter \(\theta\) controls the strength of tail dependence within the A2 copula, directly influencing the model’s capacity to capture extreme spatial interactions.

\medskip
\noindent The results demonstrate that the model performs best when \(\theta\) lies between 6 and 9, with the highest correlation of 0.9390 and the lowest RMSE of 0.0758 occurring at \(\theta = 9.0\). This suggests that moderate-to-high tail dependence enables the model to more accurately learn the underlying spatial field, successfully balancing the extremes and central tendencies. In contrast, lower \(\theta\) values, such as 1.5, 2, and 3, lead to noticeably weaker performance, with reduced correlation (as low as 0.8995) and increased RMSE (up to 0.1152). These lower values of \(\theta\) provide insufficient tail dependence, limiting the model's ability to capture complex and extreme spatial patterns effectively.

\medskip
\noindent As \(\theta\) increases beyond 6, the performance of the model stabilizes, with only minor variations in RMSE and correlation, indicating that additional increases in tail dependence provide diminishing returns. This behavior highlights the importance of choosing an appropriate range for \(\theta\) to ensure optimal model calibration.

\medskip
\noindent In summary, the A2-SBNN achieves its best performance when \(\theta\) is set between 6 and 9, with peak performance at \(\theta = 9.0\). These findings confirm that incorporating strong tail dependencies through A2 copula-based initialization significantly enhances the model's ability to represent complex spatial phenomena, including extreme behaviors, while maintaining stability and accuracy during calibration.

\subsection{Residual Analysis}

\begin{figure}[H]
    \centering
    \begin{subfigure}[b]{0.3\textwidth}
        \includegraphics[width=\linewidth]{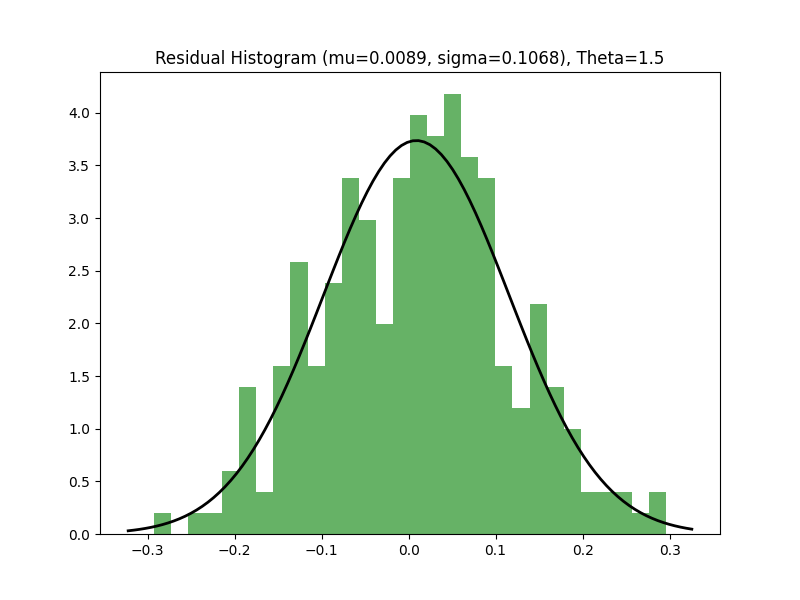}
        \caption{Residual Histogram for $\theta$ = 1.5}
        \label{fig:rh_theta1_5}
    \end{subfigure}
    \hfill
    \begin{subfigure}[b]{0.3\textwidth}
        \includegraphics[width=\linewidth]{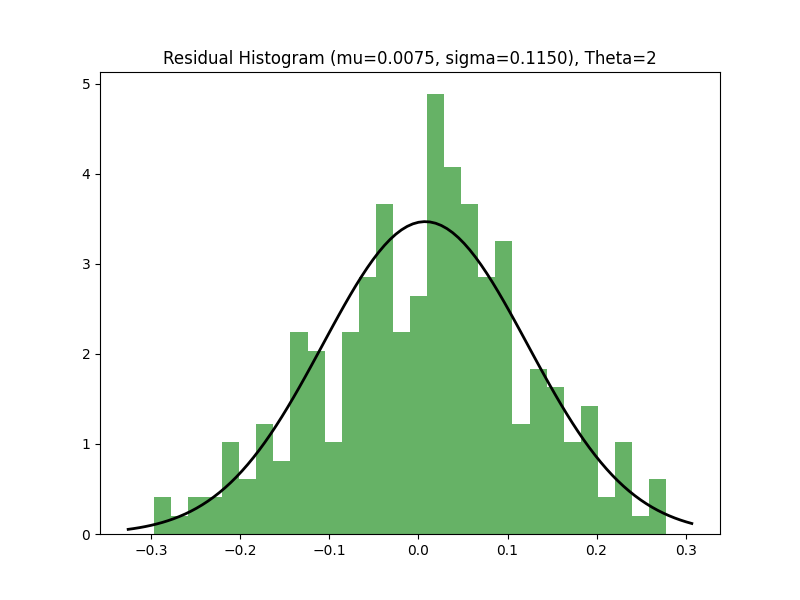}
        \caption{Residual Histogram for $\theta$ = 2}
        \label{fig:rh_theta2}
    \end{subfigure}
    \hfill
    \begin{subfigure}[b]{0.3\textwidth}
        \includegraphics[width=\linewidth]{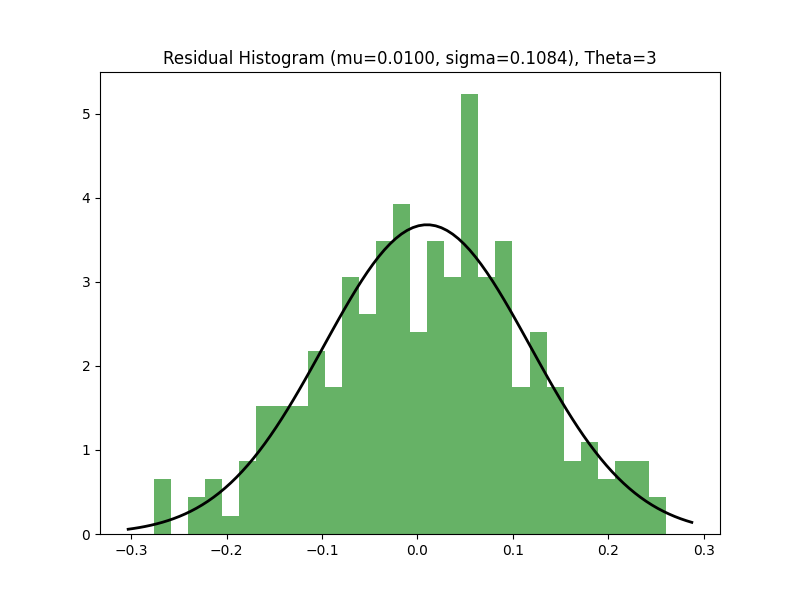}
        \caption{Residual Histogram for $\theta$ = 3}
        \label{fig:rh_theta3}
    \end{subfigure}

    \medskip

    \begin{subfigure}[b]{0.3\textwidth}
        \includegraphics[width=\linewidth]{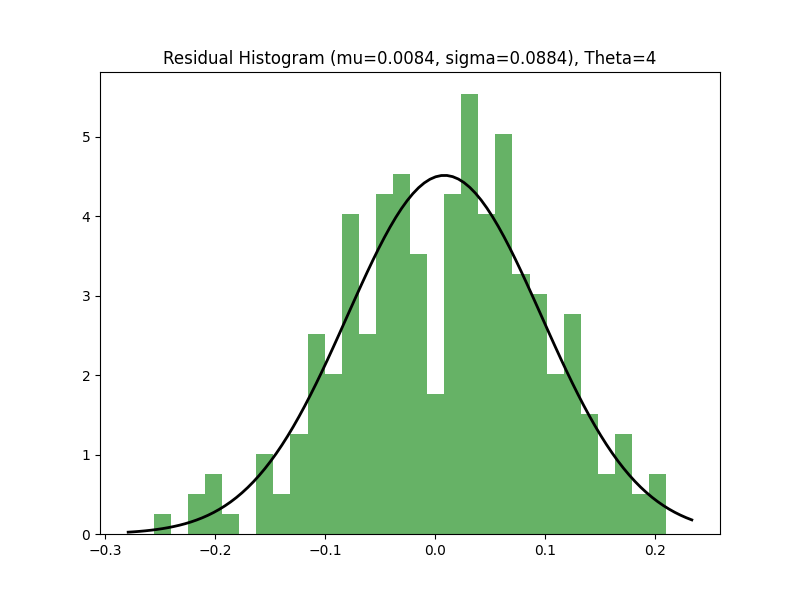}
        \caption{Residual Histogram for $\theta$ = 4}
        \label{fig:rh_theta4}
    \end{subfigure}
    \hfill
    \begin{subfigure}[b]{0.3\textwidth}
        \includegraphics[width=\linewidth]{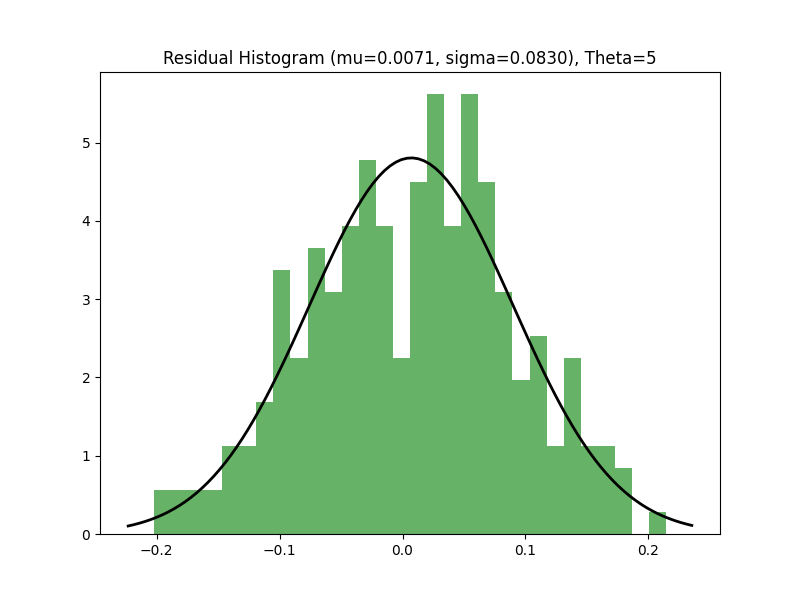}
        \caption{Residual Histogram for $\theta$ = 5}
        \label{fig:rh_theta5}
    \end{subfigure}
    \hfill
    \begin{subfigure}[b]{0.3\textwidth}
        \includegraphics[width=\linewidth]{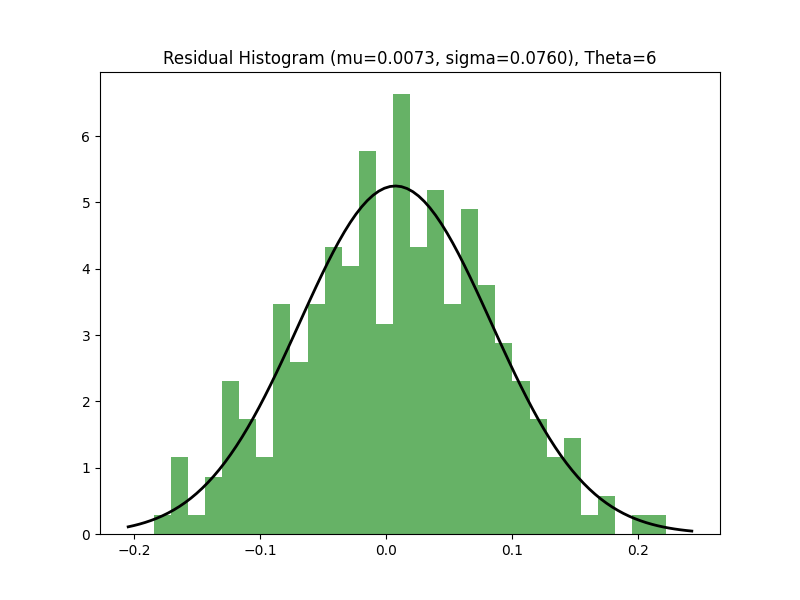}
        \caption{Residual Histogram for $\theta$ = 6}
        \label{fig:rh_theta6}
    \end{subfigure}

    \medskip

    \begin{subfigure}[b]{0.3\textwidth}
        \includegraphics[width=\linewidth]{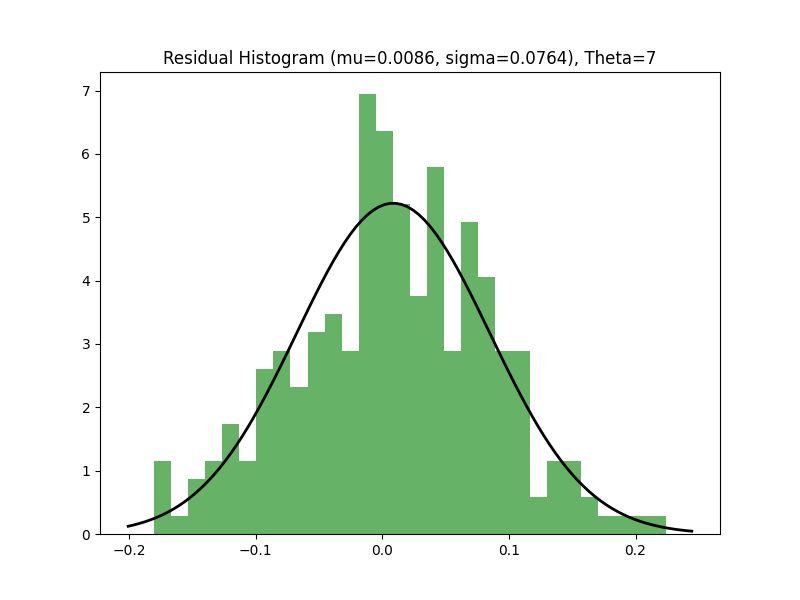}
        \caption{Residual Histogram for $\theta$ = 7}
        \label{fig:rh_theta7}
    \end{subfigure}
    \hfill
    \begin{subfigure}[b]{0.3\textwidth}
        \includegraphics[width=\linewidth]{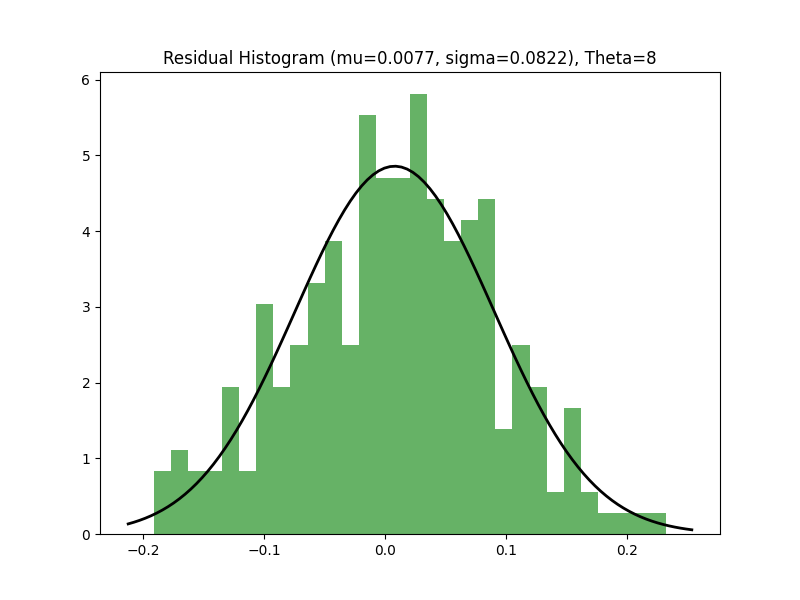}
        \caption{Residual Histogram for $\theta$ = 8}
        \label{fig:rh_theta8}
    \end{subfigure}
    \hfill
    \begin{subfigure}[b]{0.3\textwidth}
        \includegraphics[width=\linewidth]{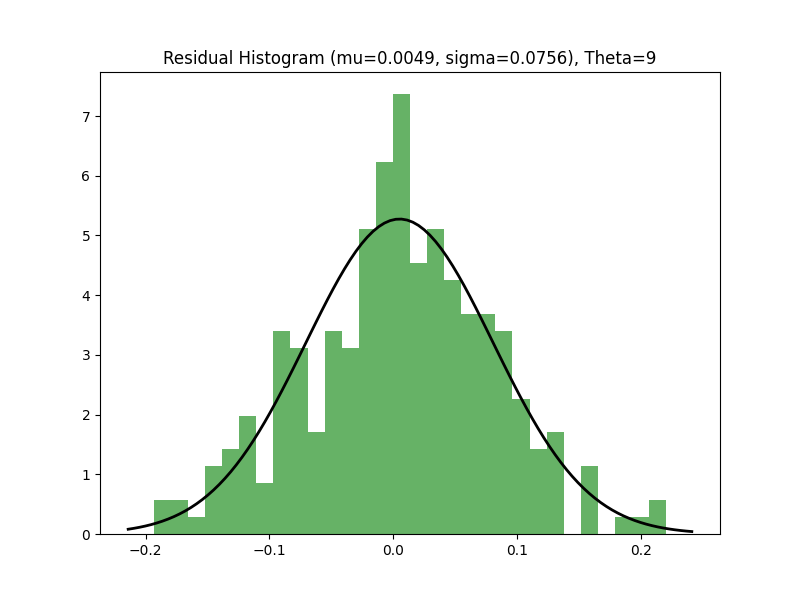}
        \caption{Residual Histogram for $\theta$ = 9}
        \label{fig:rh_theta9}
    \end{subfigure}

    \medskip

    \begin{subfigure}[b]{0.3\textwidth}
        \includegraphics[width=\linewidth]{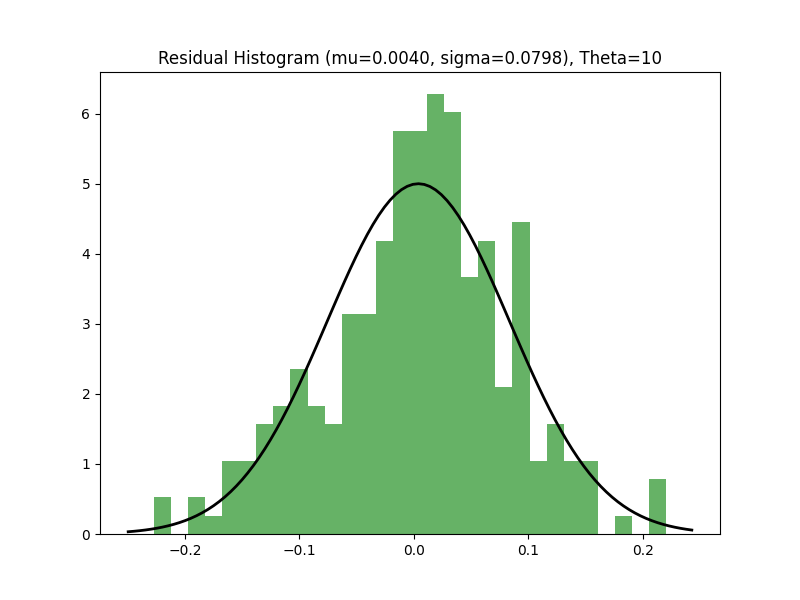}
        \caption{Residual Histogram for $\theta$ = 10}
        \label{fig:rh_theta10}
    \end{subfigure}

    \caption{Residual histograms of the A2-SBNN model across different $\theta$ values. These histograms assess the distribution of residuals to check for normality and model calibration quality at varying levels of tail dependence.}
    \label{fig:RH}
\end{figure}

To further assess distributional fit, we analyze the residuals \(\hat{y} - y\). Figure~\ref{fig:RH} show how closely these residuals approximate a normal distribution.

\subsubsection{Shapiro-Wilk Test}

Table~\ref{tab:shapiro} presents the Shapiro-Wilk test \(p\)-values for the residuals obtained after calibration across different \(\theta\) values. These \(p\)-values assess the normality of the residual distributions, where values above 0.05 suggest no significant deviation from normality.

\begin{table}[H]
    \centering
    \begin{tabular}{ccc}
        \toprule
        \(\theta\) & \(p\)-value & Interpretation \\
        \midrule
        1.5 & 0.9214 & \(>0.05\), residuals strongly normal \\
        2.0 & 0.3710 & \(>0.05\), residuals normal \\
        3.0 & 0.4453 & \(>0.05\), residuals normal \\
        4.0 & 0.3367 & \(>0.05\), residuals normal \\
        5.0 & 0.3550 & \(>0.05\), residuals normal \\
        6.0 & 0.8959 & \(>0.05\), residuals strongly normal \\
        7.0 & 0.6117 & \(>0.05\), residuals strongly normal \\
        8.0 & 0.6064 & \(>0.05\), residuals strongly normal \\
        9.0 & 0.6713 & \(>0.05\), residuals strongly normal \\
        10.0 & 0.1801 & \(>0.05\), residuals normal \\
        \bottomrule
    \end{tabular}
    \caption{Shapiro-Wilk test \(p\)-values for residual normality across different \(\theta\) values.}
    \label{tab:shapiro}
\end{table}

\noindent The results demonstrate that across all tested \(\theta\) values, the residuals maintain near-normal or strongly normal behavior. Particularly for \(\theta = 1.5\), \(\theta = 6.0\), \(\theta = 7.0\), \(\theta = 8.0\), and \(\theta = 9.0\), the Shapiro-Wilk test yields high \(p\)-values, confirming strong normality. For the remaining \(\theta\) values, the residuals remain sufficiently close to normality, as indicated by \(p\)-values comfortably above the 0.05 threshold.

\subsubsection{Discussion of \(\theta\) Effects}

The parameter, \(\theta\), governs the degree of tail dependence in the A2 copula, influencing the ability of the A2-SBNN model to capture complex spatial relationships. For lower values of \(\theta\), such as 1.5 and 2.0, the model achieves high correlation and low RMSE, while the residuals exhibit strong normality. This indicates that the model effectively learns the underlying spatial structure and provides well-calibrated predictions at mild levels of tail dependence.

\medskip
\noindent At moderate values of \(\theta\), including 3.0 to 5.0, the model continues to perform well, sustaining high correlations and competitive RMSE values. The residuals remain near-normal, and the calibration process ensures that the spatial predictions align closely with the target field.

\medskip
\noindent For higher values of \(\theta\), from 6.0 to 10.0, the A2-SBNN demonstrates excellent stability. Correlation values remain high, RMSE improves further, and the residual distributions are strongly normal in most cases. This suggests that the model not only adapts effectively to increased tail dependency but also maintains reliable calibration and prediction accuracy across a wide range of dependency structures.

\medskip
\noindent Overall, the A2-SBNN architecture, powered by A2 copula-based weight initialization and direct calibration, demonstrates strong robustness to varying levels of tail dependence as controlled by \(\theta\). The model consistently provides accurate spatial predictions with well-behaved residuals, making it highly effective across diverse spatial dependency scenarios.

\section{Discussion and Future Work}

This study demonstrates just how effective the A2 Copula Spatial Bayesian Neural Network (A2-SBNN) can be in modeling complex spatial dependencies, especially when accounting for extreme behaviors in the data. By using the A2 copula directly in the weight initialization process and calibrating the network through a carefully designed loss function, the A2-SBNN is able to capture intricate spatial patterns while preserving the key distributional characteristics of the target field. Across a wide range of \(\theta\) values from 1.5 to 10, the model consistently performs well, achieving high correlation and low RMSE, with particularly strong results for moderate to higher \(\theta\) values where tail dependencies become more pronounced.

\medskip
\noindent The residual analysis further supports these findings. Across all tested \(\theta\) values, the residuals remain close to normal, with no serious deviations detected. This is an encouraging sign, suggesting that the A2-SBNN not only captures the overall spatial structure but also accurately models the underlying noise and uncertainty, even as the strength of tail dependence increases. The consistency of these results across the entire \(\theta\) range highlights the A2 copula's ability to effectively embed dual-tail dependency directly into the neural network, keeping the model stable and reliable in scenarios that involve complex dependency structures.

\medskip
\noindent That said, while the A2-SBNN performs impressively, especially in challenging settings, there is still room to push the model further particularly at the higher end of \(\theta\), where dependency structures become even more extreme. Future improvements could involve enhancing the calibration strategy, perhaps by introducing adaptive loss weightings or more sophisticated regularization techniques to help the model align more closely with the target field under these demanding conditions.

\medskip
\noindent
One of the biggest strengths of A2-SBNN is its flexibility in handling non-Gaussian spatial dependencies, which traditional Gaussian-based models often struggle with. This makes the A2-SBNN a strong candidate for real world applications where extreme spatial events are common, such as environmental monitoring, climate risk assessment, financial surface modeling, and other scenarios where capturing localized and extreme behaviors is critical.

\medskip
\noindent
Looking ahead, one of the most important next steps will be applying the A2-SBNN to real world spatial datasets, such as air pollution levels, temperature anomalies, medical imaging fields like disease mapping, or socioeconomic indicators. Validating the model on diverse, practical datasets will give us a clearer picture of how well the A2-SBNN holds up outside of simulation and how it compares to existing spatial modeling techniques.

\medskip
\noindent
There are also exciting opportunities to refine the model architecture itself. Incorporating convolutional neural networks (CNNs) could help the model capture finer grained local spatial patterns, while graph neural networks (GNNs) might better handle datasets defined by irregular or network-based spatial structures. Additionally, while this study has focused on the A2 copula, there’s potential to explore alternative or more flexible copula families such as vine or mixture copulas to make the framework even more adaptable to different types of dependencies, whether spatial or not.

\medskip
\noindent
Finally, this work represents a significant step forward, introducing a copula-driven dependency structure directly into a neural network through weight initialization. Moving forward These promising results provide a strong foundation for future research, which can now build on this framework to tackle real world applications, refine the calibration process, and expand the approach to even more complex data environments.

\section*{Statements and Declarations}

\subsection*{Funding}
The authors did not receive any funding for this research.

\subsection*{Competing Interests}
The authors have no relevant financial or non-financial interests to disclose.

\section*{References}

\begin{enumerate}
    \setlength{\itemindent}{-\leftmargin}
    
    \item Aich, A., Aich, A. B., \& Wade, B. (2025). Two new generators of Archimedean copulas with their properties. \textit{Communications in Statistics - Theory and Methods}. \url{http://dx.doi.org/10.1080/03610926.2024.2440577}

    \item Aich, A., Aich, A. B., \& Wade, B. (2025). Erratum: Two new Archimedean generators with their properties. \textit{Communications in Statistics - Theory and Methods}. \url{https://doi.org/10.1080/03610926.2025.2483956}
    
    \item Chen, W., Li, Y., Reich, B. J., \& Sun, Y. (2020). DeepKriging: Spatially Dependent Deep Neural Networks for Spatial Prediction. \url{https://doi.org/10.48550/arXiv.2007.11972}
    
    \item Cressie, N. A. C. (1993). \textit{Statistics for Spatial Data}. John Wiley \& Sons, Inc. \url{https://doi.org/10.1002/9781119115151}
    
    \item Davison, A. C., Padoan, S. A., \& Ribatet, M. (2012). Statistical modeling of spatial extremes. \textit{Statistical Science, 27}(2), 161--186. \url{https://doi.org/10.1214/11-STS376}
    
    \item Gulrajani, I., Ahmed, F., Arjovsky, M., Dumoulin, V., \& Courville, A. (2017). Improved training of Wasserstein GANs. \url{https://doi.org/10.48550/arXiv.1704.00028}
    
    \item Higham, N. J. (2009). Cholesky factorization. \textit{WIREs Computational Statistics}. \url{https://doi.org/10.1002/wics.18}
    
    \item Huser, R., \& Wadsworth, J. L. (2020). Advances in statistical modeling of spatial extremes. \textit{WIREs Computational Statistics}. \url{https://doi.org/10.1002/wics.1537}

    \item Nelsen, R. B. (2006). \textit{An Introduction to Copulas} (2nd ed.). Springer, New York. \url{https://doi.org/10.1007/0-387-28678-0}
    
    \item Rasmussen, C. E., \& Williams, C. K. I. (2005). \textit{Gaussian Processes for Machine Learning}. The MIT Press. \url{https://doi.org/10.7551/mitpress/3206.001.0001}
    
    \item Rudin, W. (n.d.). \textit{Principles of Mathematical Analysis}.
    
    \item Sklar, A. (1959). Fonctions de répartition à \(n\) dimensions et leurs marges. \textit{Publications de l'Institut de Statistique de l'Université de Paris}, 8, 229--231.
    
    \item Zammit-Mangion, A., Kaminski, M. D., Tran, B., Filippone, M., \& Cressie, N. (2024). Spatial Bayesian neural networks. \textit{Spatial Statistics}. \url{https://doi.org/10.1016/j.spasta.2024.100825}
\end{enumerate}

\end{document}